\documentclass[12pt]{article}

\begin{document}

\begin{center}
{\Large \bf Corvino-Schoen theorem and supertranslations at spatial infinity}

\vspace{25pt}

{\large  M. \textsc{Henneaux}$^{1,2}$}

\vspace{25pt}

$^1$Physique Th\'eorique et Math\'ematique \& International Solvay Institutes,\\
Universit\'e Libre de Bruxelles, Campus Plaine C.P. 231, B-1050 Bruxelles, Belgium; \\
$^2$Coll\`ege de France, 11 place Marcelin Berthelot, 75005 Paris, France

\end{center}
\vspace{20pt}

\begin{abstract}
It is shown how the gluing theorem due to Corvino and Shoen can be naturally extended to accommodate supertranslations at spatial infinity. Logarithmic supertranslations play an intersting role in the construction.
\end{abstract}




\section{Introduction\label{ra_sec1}}
This paper is dedicated to the memory of Lars Brink, invaluable colleague and friend, who sadly passed away on October 29, 2022. Lars was not only a remarkable scientist but also a ``physicist of the world'' who cared about the development of many scientific institutions worldwide, including the International Solvay Institutes which owe him a great debt.   

The article is devoted to the BMS symmetry \cite{Bondi:1962px,Sachs:1962wk,Sachs:1962zza}, a subject to which Lars turned in his later work by exploring its light-cone formulation \cite{Ananth:2020ngt,Ananth:2020ojp}. 

One outstanding result on the initial value problem of general relativity is the ``Corvino-Schoen'' theorem \cite{Corvino:2000,Corvino:2003sp,Chrusciel:2002vb,Chrusciel:2003sr}, which enables one to consistenly glue arbitrary asymptotically flat initial data (in a sense to be made precise below) to an appropriately boosted, rotated and translated Kerr solution outside a sufficiently big volume.  The theorem can be generalized to higher dimensions  \cite{Hollands:2016oma}.

The fact that one can glue arbitrary asymptotically flat initial data  to a boosted, rotated and translated Kerr solution has sometimes be used to argue that the natural asymptotic symmetry group at spatial infinity is that of the boosted, rotated and translated Kerr family, namely,  the Poincar\'e group, with no room for supertranslations.

However, the Corvino-Schoen theorem considers from the outset asymptotic conditions involving strict parity restrictions \cite{Regge:1974zd} that freeze the possibility to perform non-trivial BMS supertranslations.   The absence of supertranslations in the asymptotic family of solutions to which one glues the supertranslation-free initial data is therefore built-in and not a surprise.

It is shown in this paper that there is a direct extension of the Corvino-Schoen theorem which naturally accommodates supertranslations.  This is achieved by considering the wider class of boundary conditions given in references \cite{Henneaux:2018hdj,Henneaux:2019yax,Fuentealba:2022xsz} and a wider class of reference solutions to which one glues the initial data.  This wider class contains a representative for each set of values of the BMS charges (Poincar\'e {\em and} supertranslation charges) and allows the possibility to perform arbitrary supertranslations.

\section{The Corvino-Schoen theorem in its original formulation}

\subsection{Boosted Kerr metrics}

The paper \cite{Corvino:2003sp} adopts the following boundary conditions
\begin{equation}
h_{ij} \equiv g_{ij} - \delta_{ij} = \frac{\bar h_{ij}(\mathbf{n}^k)}{r} + \mathcal O\left(\frac{1}{r^2}\right), \quad \bar h_{ij} (-\mathbf{n}^k) = \bar h_{ij} (\mathbf{n}^k) \label{eq:BC0a}
\end{equation}
and
\begin{equation}
\pi^{ij} = \frac{\bar \pi^{ij}(\mathbf{n}^k)}{r^2} + \mathcal O\left(\frac{1}{r^3}\right), \quad \bar \pi^{ij} (-\mathbf{n}^k) = - \bar \pi^{ij} (\mathbf{n}^k). \label{eq:BC1a}
\end{equation}
for the metric and its conjugate momentum.  Here, $\mathbf{n}^k$, which are the components of the unit normal to the sphere at infinity, parametrize the points on the sphere.  One sometimes equivalently writes $f (\mathbf{n}^k)\equiv f(\theta, \varphi)$. The leading orders in the asymptotic expansions obey strict parity conditions, but the subleading ones are not subject to such requirements.  These boundary conditions were proposed in the article \cite{Regge:1974zd}, the parity conditions on the leading order being imposed in order to make all the Poincar\'e charges $(M_{\mu \nu}, P_\mu)$ finite. They were adopted in \cite{Corvino:2003sp,Chrusciel:2003sr} for that reason.   We call the boundary conditions (\ref{eq:BC0a}) and (\ref{eq:BC1a}) ``strict parity conditions'',  with the understanding throughout that the strict parity conditions are actually only imposed on the leading order components.

Although not stressed in those works, but equally importantly,  the boundary conditions (\ref{eq:BC0a}) and (\ref{eq:BC1a}) also make the action and the symplectic form finite \cite{Henneaux:2018hdj,Henneaux:2019yax} (see also \cite{Beig:1987zz}).

The reference family of solutions to which one glues initial data fulfilling (\ref{eq:BC0a}) and (\ref{eq:BC1a}) must be such that there is one member of the family for any set of values of the Poincar\'e charges $(M_{\mu \nu}, P_\mu)$. It is natural to adopt the Kerr family obtained by boosting, rotating and translating the standard Kerr metrics with arbitrary mass and angular momentum.  We call for simplicity these Poincar\'e-transformed metrics just ``the boosted Kerr solutions''. We denote the corresponding ``boosted Kerr'' initial data by $g_{ij \; (m,a,P)}^{\textrm{Kerr}}$ and $\pi^{ij \; \textrm{Kerr}}_{(m,a,P)}$, where $P\equiv({\Lambda^\mu}_\nu, a^\mu)$ is the Poincar\'e transformation used to boost, rotate and translate the Kerr solutions.  The initial data of the boosted Kerr family are easily verified to fulfill the boundary conditions (\ref{eq:BC0a}) and (\ref{eq:BC1a}) as well as the constraint equations.

One covers in this way all possible values of the Poincar\'e charges. There is some redundancy in the parametrization since the subgroup of rotations along the $z$-axis and time translations leaves the Kerr metric invariant.  One can eliminate this redundacy by appropriately restricting the Poincar\'e group element used in rotating  and translating  the solutions before acting with a pure boost (one restricts the axis of rotation to be orthogonal to the $z$ axis and considers  spatial translations only).

\subsection{Gluing theorem (Corvino-Schoen)}

The Corvino-Schoen theorem states: 

\subsubsection*{Theorem}
Let $(g_{ij}, \pi^{ij})$ be a solution of the constraint equations on $R^3$ which obeys the asymptotic conditions (\ref{eq:BC0a}) and (\ref{eq:BC1a}).  Then there is a $\rho>0$ and a solution $(\bar g_{ij}, \bar \pi^{ij})$  of the constraint equations such that $(\bar g_{ij}, \bar \pi^{ij})$ coincides with $(g_{ij}, \pi^{ij})$ for $r \leq \rho$ and agrees with a suitably chosen member of the boosted Kerr family for $r \geq 2 \rho$.  Furthermore, the Poincar\'e charges of $(g_{ij}, \pi^{ij})$ and  $(\bar g_{ij}, \bar \pi^{ij})$ can be made arbitrarily close by chosing $\rho$ big enough.

\subsubsection*{Comments}
\begin{itemize}
\item It is well known that generic initial data decaying as $1/r$ for the metric and as $1/r^2$ for the conjugate momenta (with no parity condition) will in general develop polylogarithmic terms in the Weyl curvarture tensor at null infinity \cite{Christodoulou:2000,Friedrich:2017cjg}, invalidating the peeling theorem in its strong (original) version. The parity conditions on the leading orders eliminate the leading logarithmic terms but subleading polylogarithmic terms  will in general be present. Exceptions to this behaviour are given by the Kerr metrics, which do not develop logarithmic terms at all.  This property combined with the gluing theorem has enabled the authors of  \cite{Chrusciel:2002vb} to construct a large class of solutions with a smooth null infinity.
\item While the solutions obtained by gluing form a dense subset, they are unstable concerning the development of polylogarithmic terms at null infinity, in the sense that an arbitrarily small perturbation of $(\bar g_{ij}, \bar \pi^{ij})$ obeying only the asymptotic conditions will in general develop such terms.   Polylogarithmic terms are unavoidable unless one restricts the initial data to a very specific subset \cite{ValienteKroon:2003ix}.  Arbitrary long-wavelength configurations with non-zero energy will generically lead to polylogarithmic terms, however small (but not zero) the energy is.
\end{itemize}

\section{Taking into account the supertranstranlations}

\subsection{Need to include the supertranslations}

One interesting feature of the description of the boosted Kerr family is that it involves explicitly the group element needed to generate non-vanishing values of the Poincar\'e charges. These appear as ``large diffeomorphisms'', i.e., as diffeomorphisms not vanishing at infinity. Because they change the Poincar\'e charges, these diffeomorphisms have a physical impact and should by no means be quotientized out.  It would be wrong to impose asymptotic conditions on the gravitational variables that would force the system to be in its rest frame, allowing rotations only around the z-axis and putting the origin at the center of mass.  

Among other limitations, such a drastic reduction would prevent one from considering simultaneously two isolated subsystems with non-zero relative motion and angular momenta in different directions since these subsystems cannot then both fulfill these very restricted asymptotic conditions. [These subsystems might furthermore interact for a finite time before and after which asymptotic developments in $1/r$ are relevant, with different values of the individual ``conserved charges'' for each subsystem before and after the interaction.]

One could imagine formulating a version of the Corvino-Schoen theorem for the restricted set of boundary conditions just alluded to,  which force the conserved charges to be zero, except $P^0 = m$ and $J^z = a$ (the Casimirs).  The invariance group of these boundary conditions is the two-dimensional subalgebra $R \times SO(2)$. Although mathematically consistent, the existence of such a restricted version of the theorem could not be used to infer that the Poincar\'e group is absent at spatial infinity.  This would miss important physics. The full Poincar\'e group is physically relevant.

This observation sheds light on the question of supertranslations at spatial infinity.  The original version of the Corvino-Schoen theorem does not allow supertranslations, but this does not mean that supertranslations are irrelevant at spatial infinity. They have simply not been included.  Just as the Poincar\'e transformations, they have, however, a non trivial action on the physical states.  It would be a limitation to freeze the possibility to perform them, in much the same way as freezing the possibility to perform spatial translations or rotations about arbitrary axes does not allow the most general (and useful) physical transformation of the system.

\subsection{Hamiltonian description of the supertranslations - twisted boundary conditions}

It is easy to take the supertranslations into account by the same group theoretical procedure of acting with supertranslations on the above solutions.  To achieve this task, one needs to know how supertranslations act at spatial infinity.  This was worked out in references \cite{Henneaux:2018hdj,Henneaux:2019yax}.  It was found that supertranslations ``twist'' the strict parity conditions of reference \cite{Regge:1974zd} by an $\mathcal O(1)$-diffeomorphism taking a specific form.  The new boundary conditions read
\begin{equation}
h_{ij} \equiv g_{ij} - \delta_{ij} = \frac{\bar h_{ij}(\mathbf{n}^k)}{r} + U_{ij} + \mathcal O\left(\frac{1}{r^2}\right) , \quad \bar h_{ij} (-\mathbf{n}^k) = \bar h_{ij} (\mathbf{n}^k) \label{eq:BC0b}
\end{equation}
with
\begin{equation}
 U_{ij} = 2\partial_i \partial_j (rU) ,  \quad  U = U (\mathbf{n}^k) = O(1), \quad   U (-\mathbf{n}^k) = - U (\mathbf{n}^k), 
\end{equation}
which implies 
\begin{equation}
U_{ij} =  O(\frac{1}{r}), \quad U_{ij}  (-\mathbf{n}^k) = -U_{ij}  (\mathbf{n}^k).
\end{equation}
Similarly,
\begin{equation}
\pi^{ij} = \frac{\bar \pi^{ij}(\mathbf{n}^k)}{r^2} + V^{ij} + \mathcal O\left(\frac{1}{r^3}\right), \quad \bar \pi^{ij} (-\mathbf{n}^k) = - \bar \pi^{ij} (\mathbf{n}^k),  \label{eq:BC1b}
\end{equation}
with
\begin{equation}
V^{ij} = \partial^i \partial^j V - \delta^{ij}   \partial^k \partial_k V, \quad  V = \mathcal{O}(1), \quad V (-\mathbf{n}^k) = V (\mathbf{n}^k)
\end{equation}
which implies 
\begin{equation}
 V^{ij} = \mathcal O\left(\frac{1}{r^2}\right), \quad V^{ij}  (-\mathbf{n}^k) = V^{ij}  (\mathbf{n}^k)
\end{equation}
It should be noted that the respective $\ell=0$ and $\ell=1$ spherical harmonics components of $V$ and $U$ drop from these formulas, which describe ``pure supertranslations''.  Ordinary translations in time ($V_{\ell=0}$) and in space ($U_{\ell = 1}$) do not twist the parity conditions.

One might be worried that the functions $U$ and $V$ are not smooth at the origin $r=0$.  However, one should recall that only the asymptotic behaviour of the diffeomorphisms $(\zeta^\perp = V + \textrm{``subleading terms''}, \zeta^i = \partial^i (rU))+ \textrm{``subleading terms''})$ associated with them is relevant for the asymptotic symmetry considerations.  One can extend these diffeomorphisms ``inside'' in such a way that they are smooth.  Because these are diffeomorphisms, they preserve the constraints (provided, of course, that one takes the full (non-linear) expressions for the variations of the fields in the bulk).

We also note that because the extra terms in the boundary conditions originate from diffeomorphisms (which linearize to leading order), they will not change the leading behaviour of the Weyl tensor.  In particular, the leading term involving a logarithm  in the polylogarithmic expansion at null infinity will be absent, just as it is absent with the original boundary conditions (\ref{eq:BC0a}) and (\ref{eq:BC1a}).

\subsection{Action of supertranslations on fields $U$ and $V$ - BMS charges}
The action of a supertranslation on the canonical fields (\ref{eq:BC0b}) and (\ref{eq:BC1b}) is simply to shift $U$ and $V$ as \cite{Henneaux:2018hdj,Henneaux:2019yax}
\begin{equation}
U \rightarrow U + W, \qquad V \rightarrow V + T  \label{Eq:Shifts}
\end{equation}
where $W(\mathbf{n}^k)$ and $T(\mathbf{n}^k)$ are respectively odd and even functions of the angles.  The action is abelian to this leading order.

We recall that the supertranslations, parametrized by a  function of the angles $\alpha (\mathbf{n}^k)$ at null infinity, are equivalently parametrized by an odd function $W(\mathbf{n}^k)$ and an even function $T(\mathbf{n}^k)$ of the angles in the Hamiltonian description.  The explicit change of Lie algebra bases is given in references \cite{Troessaert:2017jcm,Henneaux:2018cst}.
[Note, however, that the second of these references adopts different boundary conditions on the canonical variables, but the supertranslations take the same form.]

It was proved in references \cite{Henneaux:2018hdj,Henneaux:2019yax} that the extended boundary conditions  (\ref{eq:BC0b}) and (\ref{eq:BC1b}) keep the finiteness of the action if one imposes that the constraints, which would decay as $r^{-3}$ at infinity, actually decrease two orders of $r$ faster (if one also requests that relevant integrals involving the constraints are finite off the constraint surface).  Furthermore, Poincar\'e invariance is maintained and the corresponding genuine charge-generators are all finite.  The supertranslation charges can also be computed by standard Hamiltonian methods -- the explicit Hamiltonian expression is given in references \cite{Henneaux:2018hdj,Henneaux:2019yax} -- and close in the Poisson bracket according to the BMS algebra.  They match the past limit ($u \rightarrow - \infty$) of the BMS charges defined on Scri$^+$ \cite{Troessaert:2017jcm}.

\subsection{Physical meaning of $U$ and $V$}
There is another good reason to keep the surface fields $U$ and $V$ in the formalism and not set them equal to zero.   These fields have a direct physical interpretation.  Indeed, they combine to form the ``Goldstone boson of spontaneously broken supertranslation invariance'' (also called ``boundary graviton'') $C$  introduced in \cite{Strominger:2013jfa,Strominger:2017zoo} at null infinity, which plays an important role in the extraction of the physical angular momentum flux at infinity \cite{Javadinezhad:2022ldc}. This field can be shifted by the passage of gravitational wave \cite{Strominger:2014pwa}, the so-called ``memory effect''.

The explicit match between $U$, $V$ and $C$ can be made at the past of future null infinity \cite{Fuentealba:2023syb} ($\lim_{u \rightarrow - \infty} C$).  

\subsection{Supertranslation charges and strict parity conditions}

It is interesting to note that the supertranslation charges of the configurations obeying strict parity conditions are not necessarily zero.  This can be verified by direct inspection of the BMS charges. One way to understand this phenomenon is through the BMS algebra itself.  By boosting a solution obeying strict parity conditions  with non-vanishing spatial momentum, one can generate a non-vanishing supertranslation charge while remaining within the class of solutions obeying strict parity conditions.

\section{Logarithmic supertranslations}

\subsection{How to ``supertranslation-charge'' a solution}

If one acts with a supertranslation on a solution with $U=0, V=0$ and vanishing supertranslation charges, one gets a solution with $U \not=0$ and $V \not=0$.  However, this solution still has vanishing supertranslation charges, since the supertranslations form an abelian subalgebra and their charge-generators form a representation of it with no central term \cite{Henneaux:2018hdj,Henneaux:2019yax}.  To ``supertranslation-charge'' a solution, one needs to proceed differently. 

It is true, as we have recalled, that by boosting a solution with non-vanishing spatial momentum, one can generate a non-vanishing supertranslation charge.  Which values of the supertranslation charges  one can reach in this mannner is an issue to which one does not actually need the answer because there is a simpler way to ``supertranslation-charge'' a solution: one simply performs logarithmic supertranslations.

Logarithmic supertranslations have been introduced in the Hamiltonian context in reference \cite{Fuentealba:2022xsz}. They correspond to diffeomorphisms that grow up logarithmically in $r$, with coefficients that involve two functions $U^{\textrm{log}}, V^{\textrm{log}}$ of the angles, one even and one odd, with  the $\ell = 0$ and $\ell=1$ modes absent (pure gauge).  When these transformations are included, the boundary conditions are further enlarged and read
\begin{equation}
h_{ij} = g_{ij}- \delta_{ij} =  \Delta_{ij}^{\textrm{log}} + U_{ij} + h_{ij}^{\textrm{RT}} \label{eq:BC0c}
\end{equation}
and
\begin{equation}
\pi^{ij} =  \Gamma^{ij}_{\textrm{log}} + V^{ij} + \pi^{ij}_{\textrm{RT}} \label{eq:BC1c}
\end{equation}
where  $\Delta_{ij}^{\textrm{log}}$ and $\Gamma^{ij}_{\textrm{log}}$ are the changes in the metric and its conjugate momentum due to the logarithmic supertranslations.   The terms $h_{ij}^{\textrm{RT}}$ and $\pi^{ij}_{\textrm{RT}}$ obey the strict parity conditions on the leading orders given in equations (\ref{eq:BC0a}) and (\ref{eq:BC1a}).  The terms $\Delta_{ij}^{\textrm{log}}$ and $\Gamma^{ij}_{\textrm{log}}$  are of respective orders  $\mathcal{O}(\frac{\ln r}{r})$and $\mathcal{O}(\frac{\ln r}{r^2})$ and provide an additional twist to these strict parity conditions. Their explicit expressions in terms of $U^{\textrm{log}}, V^{\textrm{log}}$ can be found in reference \cite{Fuentealba:2022xsz}.  Because logarithmic supertranslations are diffeomorphisms, they do not change the asymptotic behaviour of the Weyl tensor components to leading order and hence will not introduce leading polylogarithmic singularities at null infinity with respect to the boundary conditions (\ref{eq:BC0a}) and (\ref{eq:BC1a}).  
 
What is crucial for our current purposes is that the logarithmic supertranslation charges are canonically conjugate to the pure supertranslations charges.  These form together an abelian algebra, which is centrally extended in their canonical realization.   By acting with logarithmic supertranslations, one can thus shift at will the pure supertranslation charges.  And similarly, by acting with supertranslations, one can shift at will the logarithmic supertranslation charges.  The charge-generators of the logarithmic supertranslations turn out to be $U$ and $V$ so that this shift property is the expression of (\ref{Eq:Shifts}). 

We will also adopt the redefinition of reference \cite{Fuentealba:2022xsz} of the Hamiltonian symmetry generators, which yields Poincar\'e generators (including the homogeneous ones) that commute with (i.e., do not transform under) the pure supertranslations and the logarithmic supertranslations. 

\subsection{Generalized Kerr metrics}
One can extend the boosted Kerr family of solutions to the constraint equations in the same spirit as one extends the original Kerr family by acting with the Poincar\'e group.  One just acts also with both logarithmic and BMS supertranslations.

We call these transformed metrics ``the generalized Kerr solutions''. We denote the corresponding ``generalized Kerr initial data'' by $g_{ij \; (m,a,P, U,V, U^{\textrm{log}}, V^{\textrm{log}})}^{\textrm{Kerr}}$ and $\pi^{ij \; \textrm{Kerr}}_{(m,a,P,U,V,U^{\textrm{log}}, V^{\textrm{log}})}$ where $P\equiv({\Lambda^\mu}_\nu, a^\mu)$ is the Poincar\'e transformation used to boost, rotate and translate the Kerr solutions and $(U,V, U^{\textrm{log}}, V^{\textrm{log}})$ the pure supertranslation and the logarithmic supertranslation with which one acts after these Poincar\'e transformations have been performed.  Not only do the initial data of the supertransled boosted Kerr family fulfill the boundary conditions (\ref{eq:BC0c}) and (\ref{eq:BC1c}) and the constraint equations, but they also cover by construction all possible values of the supertranslation charges.

\subsection{Supertranslation-compatible reformulation of the Corvino-Schoen theorem}

The supertranslation-compatible reformulation of the Corvino-Schoen theorem is now quite obvious:

\subsubsection*{Theorem}
Let $(g_{ij}, \pi^{ij})$ be a solution of the constraint equations on $R^3$ which obeys the asymptotic conditions (\ref{eq:BC0c}) and (\ref{eq:BC1c}).  Then there is a $\rho>0$ and a solution $(\bar g_{ij}, \bar \pi^{ij})$  of the constraint equations such that $(\bar g_{ij}, \bar \pi^{ij})$ coincides with $(g_{ij}, \pi^{ij})$ for $r \leq \rho$ and agrees with a suitably chosen member of the generalized Kerr family for $r \geq 2 \rho$.  Furthermore, the Poincar\'e charges and the logarithmic and pure supertranslation charges of $(g_{ij}, \pi^{ij})$ and  $(\bar g_{ij}, \bar \pi^{ij})$ can be made arbitrarily close by chosing $\rho$ big enough.

\subsubsection*{Proof}
Let  $(g_{ij}, \pi^{ij})$ be a solution of the constraint equations that obeys the asymptotic conditions (\ref{eq:BC0c}) and (\ref{eq:BC1c}).  By applying logarithmic and pure supertranslations, one can eliminate the twist, i.e., set  $ \Delta_{ij}^{\textrm{log}}$, $U_{ij}$ $\Gamma^{ij}_{\textrm{log}}$ and  $V^{ij}$ equal to zero.  These logarithmic and pure supertranslations can be assumed to vanish in the ball $r \leq \sigma$ with $\sigma$ as large as one wants since only their asymptotic values matter. The resulting initial data obey the strict parity conditions (\ref{eq:BC0a}) and (\ref{eq:BC1a}).  One can thus apply the Corvino-Schoen gluing theorem and produce a solution $(\bar g'_{ij}, \bar \pi'^{ij})$ of the corresponding gluing problem. By undoing the supertranslations one gets then a desired solution $(\bar g''_{ij}, \bar \pi''^{ij})$ with almost all the requested properties. The Poincar\'e charges can be made to be arbitrarily close and  the logarithmic supertranslation charges ($U$ and $V$) of $(g_{ij}, \pi^{ij})$ and $(\bar g''_{ij}, \bar \pi''^{ij})$ exactly match.  But the pure supertranslation charges might not match. This is because the pure supertranslation charges of a configuration obeying strict parity conditions may not vanish in its rest frame, while those of the Kerr solution do (in the rest frame).   But these can be made to match by performing a further logarithmic supertranslation at infinity on $(\bar g''_{ij}, \bar \pi''^{ij})$ (vanishing for $r< \psi$ with $\psi$ arbitrarily large) to bring these charges to the desired value.  This does not change the other charges as the algebra indicates.

\subsection{Half-way version}

If follows from the above construction that the glued solution $(\bar g_{ij}, \bar \pi^{ij})$ may involve logarithmic diffeomorphisms at infinity even if $( g_{ij},  \pi^{ij})$ does not.  One can avoid these logarithmic diffeomorphisms and formulate a weaker, ``half-way''  form of the gluing theorem for the bounday conditions (\ref{eq:BC0b}) and (\ref{eq:BC1b}),  provided one allows a mismatch of the pure supertranslation charges.  

We define the ``half-generalized'' Kerr family the solutions obtained by acting on the Poincar\'e transformed Kerr solutions by a general pure supertranslations, with no logarithmic supertranslations.  These generate non vanishing $U$'s and $V$'s but no logarithmic terms in the metric and conjugate momenta.  We write $g_{ij \; (m,a,P, U,V)}^{\textrm{Kerr}}$ and $\pi^{ij \; \textrm{Kerr}}_{(m,a,P,U,V)}$ for the corresponding ``half-generalized'' initial data.  These solutions cover all values of the Poincar\'e charges, as well as all values of the logarithmic charges $U$ and $V$ (which cannot be interpreted as generators since the logarithmic supertranslations which they generate have been frozen).  

\subsubsection*{Theorem}
Let $(g_{ij}, \pi^{ij})$ be a solution of the constraint equations on $R^3$ which obeys the asymptotic conditions (\ref{eq:BC0b}) and (\ref{eq:BC1b}).  Then there is a $\rho>0$ and a solution $(\bar g_{ij}, \bar \pi^{ij})$  of the constraint equations such that $(\bar g_{ij}, \bar \pi^{ij})$ coincides with $(g_{ij}, \pi^{ij})$ for $r \leq \rho$ and agrees with a suitably chosen member of the half-generalized Kerr family for $r \geq 2 \rho$.  Furthermore, the Poincar\'e charges  of $(g_{ij}, \pi^{ij})$ and  $(\bar g_{ij}, \bar \pi^{ij})$ can be made arbitrarily close by chosing $\rho$ big enough.

The asymptotic coefficients of $(g_{ij}, \pi^{ij})$ and  $(\bar g_{ij}, \bar \pi^{ij})$  match, showing the equality of the logarithmic supertranslation charges.    The charges of the pure supertranslations, do, however,  differ in general.

\section{Conclusions}

We have indicated how the Corvino-Schoen theorem can be extended to encompass BMS supertranslations.  One can glue any set of intial data satisfying the boundary conditions (\ref{eq:BC0c}) and (\ref{eq:BC1c}) to a member of the generalized Kerr family with arbitrarily close log-BMS charges. The glued generalized Kerr metric involves in general logarithmic terms to leading order on the Cauchy hypersurface in order to match the charges, but these terms are of a very special nature since they entirely originate from a logarithmic diffeomorphism called ``logarithmic supertranslations''.  

It should be mentioned in that respect that  a general study of initial data with polylogarithmic terms going beyond those originating from diffeomorphisms has been given in the work \cite{Hintz:2017xxu}.  Such terms are sometimes needed at subleading order in order to solve the constraints \cite{Beig:1987zz,Huang:2010yd}. 

 We stress again that our leading logarithmic terms in the metric ($\mathcal O(\ln r /r)$) and in the conjugate momentum ($\mathcal O(\ln r /r^2)$) are induced by diffeomorphisms and do not affect invariants such as the Weyl tensor to leading order (where the equations linearize).  We stress also that the polylogarithmic terms arising at null infinity are not related to polylogarithmic terms on the Cauchy surface.  This is seen in the most transparent way in the case of the massless scalar field with smooth initial data on a Cauchy surface, which decays as $1/r$.  Even in the absence of polylogarithmic terms on the Cauchy surface, polylogarithmic terms generically develop at null infinity as explicitly computed in \cite{Henneaux:2018mgn} (and there is no gauge invariance that can be invoked to perhaps remove them there).  This is also true for electromagnetism in Minkowski space \cite{Henneaux:2018gfi,Henneaux:2019yqq}.

\section*{Acknowledgments}
It is a pleasure to thank Oscar Fuentealba and C\'edric Troessaert for useful collaborations on topics directly related to this paper. This work was partially supported  by FNRS-Belgium (conventions FRFC PDRT.1025.14 and IISN 4.4503.15), as well as by funds from the Solvay Family.

\bibliographystyle{ws-rv-van}




\end{document}